# Zero-shot 3D Segmentation of Abdominal Organs in CT Scans Using Segment Anything Model 2: Adapting Video Tracking Capabilities for 3D Medical Imaging

Yosuke Yamagishi, MD[1]*, Shouhei Hanaoka, MD, PhD[1,2], Tomohiro Kikuchi, MD, MPH, PhD[3,4], Takahiro Nakao, MD, PhD[4], Yuta Nakamura, MD, PhD[4], Yukihiro Nomura, PhD[4,5], Soichiro Miki, MD, PhD[4], Takeharu Yoshikawa, MD, PhD[4], Osamu Abe, MD, PhD[1,2]

1. Division of Radiology and Biomedical Engineering, Graduate School of Medicine, The University of Tokyo
2. Department of Radiology, The University of Tokyo Hospital
3. Department of Radiology, School of Medicine, Jichi Medical University
4. Department of Computational Diagnostic Radiology and Preventive Medicine, The University of Tokyo Hospital
5. Center for Frontier Medical Engineering, Chiba University

* yamagishi-yosuke0115@g.ecc.u-tokyo.ac.jp

## Abstract

**Objectives:**

To evaluate the zero-shot performance of Segment Anything Model 2 (SAM 2) in 3D segmentation of abdominal organs in CT scans, and to investigate the effects of prompt settings on segmentation results.

**Materials and Methods:**

In this retrospective study, we used a subset of the TotalSegmentator CT dataset from eight institutions to assess SAM 2's ability to segment eight abdominal organs. Segmentation was initiated from three different z-coordinate levels (caudal, mid, and cranial levels) of each organ. Performance was measured using the Dice similarity coefficient (DSC). We also analyzed the impact of "negative prompts," which explicitly exclude certain regions from the segmentation process, on accuracy.

**Results:**

123 patients (mean age, 60.7 ± 15.5 years; 63 men, 60 women) were evaluated. As a zero-shot approach, larger organs with clear boundaries demonstrated high segmentation performance, with mean DSCs as follows: liver 0.821 ± 0.192, right kidney 0.862 ± 0.212, left kidney 0.870 ± 0.154, and spleen 0.891 ± 0.131. Smaller organs showed lower performance: gallbladder 0.531 ± 0.291, pancreas 0.361 ± 0.197, and adrenal glands, right 0.203 ± 0.222, left 0.308 ± 0.234. The initial slice for segmentation and the use of negative prompts significantly influenced the results. By removing negative prompts from the input, the DSCs significantly decreased for six organs.

**Conclusion:**

SAM 2 demonstrated promising zero-shot performance in segmenting certain abdominal organs in CT scans, particularly larger organs. Performance was significantly influenced by input negative prompts and initial slice selection, highlighting the importance of optimizing these factors.

## Introduction

Medical image segmentation is a critical task in radiology, playing a vital role in diagnosis, treatment planning, and clinical research [1, 2]. Traditionally, this process has been labor-intensive, requiring manual delineation by skilled radiologists. However, recent advancements in deep learning have revolutionized this field, expanding the scope of automated analysis and significantly enhancing performance across diverse medical imaging tasks.

The Segment Anything Model (SAM), introduced by Meta AI, represented a significant leap forward in image segmentation technology [3]. Trained on over a billion masks, SAM demonstrated remarkable versatility in segmenting a wide array of objects across various domains. SAM's zero-shot performance - its ability to segment objects it has never seen during training - in medical images has been extensively evaluated [4, 5], and specialized models such as MedSAM [6], which underwent additional training for medical imaging applications, have been introduced. These developments have demonstrated SAM's potential in radiological domains, including CT and MRI. However, SAM was primarily designed for 2D image segmentation, which imposed inherent limitations on its direct applicability to 3D volumetric data.

SAM 2, released in July 2024, introduced video segmentation capabilities [7], applicable to 3D medical imaging like CT scans. Although not specifically designed for medical use, its zero-shot ability and video tracking features offer a promising approach to 3D medical image segmentation, potentially overcoming limitations of traditional methods that require extensive domain-specific training. Testing SAM 2's zero-shot performance is crucial because it could significantly reduce the need for large, annotated medical datasets and specialized model training, potentially accelerating the deployment of AI in various medical imaging applications.

SAM 2's zero-shot performance in radiology and the impact of input factors remain understudied, despite evaluations in surgical video segmentation [8] and specialized versions like Medical SAM2 [9]. We assess SAM 2's zero-shot performance in medical imaging, examining how target organ size, initial slice selection, and negative prompts influence its segmentation accuracy. These factors are crucial for optimizing SAM 2's performance in radiological applications.

To our knowledge, this comprehensive evaluation combining zero-shot performance assessment and input factor analysis is the first of its kind for SAM 2 applied to 3D medical imaging. Our exploration is analogous to large language models, where performance varies significantly based on prompt adjustments [10]. By examining prompt engineering in segmentation, we aim to provide deeper insights into adapting general-purpose AI models for specialized medical imaging applications.

## Materials and Methods

Since the study used an open dataset, approval from the institutional review board and patient consent were not necessary. This study was conducted as a retrospective study. This study adheres to the Checklist for Artificial Intelligence in Medical Imaging (CLAIM): 2024 Update [11].

### Dataset

We aimed to evaluate the segmentation performance of major organs within the imaging range of abdominal CT, one of the most common medical imaging modalities. To conduct this performance evaluation, we utilized a subset of the TotalSegmentator CT dataset version 1.0 [12]. The TotalSegmentator dataset is a large-scale, multi-organ segmentation dataset collected from 8 institutions. We selected this dataset for its comprehensive organ segmentation masks and institutional metadata for each case. These segmentation masks have undergone expert verification to ensure high quality, making the dataset particularly suitable for our study.

Our study included cases that encompassed the abdominal region, while CT angiography scans were excluded from the analysis.

To ensure representation from all 8 institutions while managing the dataset size, we implemented a sampling strategy. We set a maximum of 20 cases per institution and randomly selected cases up to this limit. For institutions with fewer than 20 cases, all available cases were included.

We focused on 8 major abdominal organs for our analysis:

1. Liver
2. Right Kidney
3. Left Kidney
4. Spleen
5. Gallbladder
6. Pancreas
7. Right Adrenal Gland
8. Left Adrenal Gland

These organs were chosen based on their clinical significance and visibility in standard abdominal CT scans. To account for potential annotation deficiencies, we excluded extremely small volume masks by setting a threshold of 100 voxels. Masks below this threshold were omitted from the analysis. The dataset selection flowchart is illustrated in Figure 1.

**Data Preprocessing**

The dataset was available in NIfTI file format. For SAM 2 inference, we extracted each horizontal slice from the 3D volumes to create subsets of 2D images for each scan. We applied windowing to the CT scans, using a window level of 50 and a window width of 400 Hounsfield units. Following windowing, we performed min-max scaling on the data. The scaled values were then converted to 8-bit integers, resulting in a range of 0-255. These processed 2D images were saved as sequential JPEG files.

For SAM 2 inference, we selectively processed only the slices containing abdominal organs. This approach focused on optimizing computational efficiency, resulting in faster inference speeds.

**Analysis of Organ Mask Volumes**

For the volumetric analysis, we utilized the existing segmentation masks from the TotalSegmentator dataset to calculate the volume of each organ in voxels. We chose to measure in voxels rather than physical units because our model inputs do not consider voxel scale.

Additionally, we analyzed cross-sectional areas of organ masks along the z-axis. For each organ, we calculated mean mask areas at the 25th, 50th, and 75th percentile z-coordinates, corresponding to the initial prompt locations used in SAM 2 segmentation.

**SAM 2 Implementation for 3D Medical Image Segmentation**

SAM 2 is a segmentation model not specifically designed for medical images, but rather for general videos, such as sports or animal videos. These models are trained on a large dataset, enabling them to perform segmentation on any objects. SAM 2's main feature is its ability to support not only 2D images but also videos. By inputting coordinates indicating the target object for segmentation, SAM 2 can track and segment objects appearing within the video. An overview of CT volume segmentation using SAM 2's video predictor is shown in Figure 2.

- *Adaptation for 3D Medical Imaging*

Although SAM 2 was originally intended for tracking objects in general videos, we recognized that 3D volumes like CT and MRI scans can be considered as videos composed of numerous 2D images. Utilizing publicly available datasets with segmentation masks, we applied preprocessing compatible with SAM 2's video prediction capabilities. This allowed us to construct a pipeline capable of performing multi-organ segmentation in a zero-shot manner, without additional training of SAM 2.

- *Bidirectional Prediction Approach*

While SAM 2's video prediction is unidirectional, it can process in both directions from the initial frame. To obtain a complete segmentation mask for the entire volume, we implemented a simple bidirectional approach: we performed inference in both the forward direction from the starting slice to the most cranial slice, and in the reverse direction from the starting slice to the most caudal slice. The two segmentation masks obtained from these bidirectional inferences were then merged to create a full 3D segmentation mask.

- *Model Inference and Prompt Setting*

SAM 2's video prediction requires input of both the video (numbered 2D images) and prompt (coordinates for the target object). In practice, the prompt must be manually specified by a user. However, given the need to evaluate a large number of objects, we devised an algorithm to automatically obtain prompts:

1. Z-coordinate focus: Using 25th (caudal-level), 50th (mid-level), and 75th percentiles (cranial-level) for comprehensive organ representation.
2. Random selection within organ boundaries:
    - Five positive prompts from within the segmentation mask
    - Five negative prompts were sampled from the region 2-3 voxels outside the mask boundary, excluding the immediate 1-voxel margin

This method maintains reproducibility, reduces bias from the user's prompting skill, and leverages SAM 2's capability to use both positive and negative prompts for improved accuracy.

In this study, we refer to the 3D segmentations initiated from each of these positions as caudal-approach, mid-approach, and cranial-approach, corresponding to segmentations starting from the caudal, mid, and cranial-level slices, respectively.

- *Model Version*

For our study, we selected the "sam2_hiera_large" model, due to its highest performance among the available versions. We used version 1.0 of the SAM 2 (https://github.com/facebookresearch/segment-anything-2). Our implementation was carried out using Python 3.10.12.

**Statistical Analysis**

To evaluate the model's performance, we calculated the Dice similarity coefficient (DSC) [13]. This evaluation was performed organ-wise across the dataset to provide a detailed analysis.

We compared the segmentation performance across the different approaches and with/without negative prompts. We performed three pairwise comparisons for the approaches: caudal-approach vs mid-approach, caudal-approach vs cranial-approach, and mid-approach vs cranial-approach. Additionally, we compared performance with and without negative prompts.

When considering organ volumes in detail, we calculated Spearman's correlation coefficients to examine the relationship between object volumes and DSCs [14, 15].

To account for multiple comparisons, we applied the Bonferroni correction. After Bonferroni correction, $p < 0.05/3$ (approximately 0.0167) was considered statistically significant.

All statistical analyses were conducted using SciPy version 1.14.0.

## Results

**Data Characteristics**

Our sampling strategy resulted in a total of 123 scans. Twenty scans each were selected from five institutions, while the remaining three institutions contributed 5, 5, and 13 cases respectively. The average age of the patients in our selected sample was 60.7 ± 15.5 years. The gender distribution consisted of 63 males and 60 females.

903 organ segmentations were obtained from 123 scans. 12 masks with volumes of 100 voxels or smaller were then excluded from the analysis. The final dataset consisted of 891 organ segmentations.

**Analysis of Organ Mask Volumes and Areas**

Organ volumes were detailed in Table 1, measured in voxels. The liver was the largest organ, followed by the spleen. Kidneys were the next largest, with similar volumes for left and right. Compared to the liver's mean volume, the pancreas was approximately 1/25, the gallbladder less than 1/70, and both adrenal glands less than 1/400 in size. These latter organs (pancreas, gallbladder, and adrenal glands) can be categorized as small organs.

Organ cross-sectional area analysis showed diverse trends across 8 organs. The liver was largest, increasing caudally to cranially. The pancreas steadily increased. Adrenal glands, though smallest, peaked at mid-level. Details in Supplemental Figure 1.

**Multi-organ Segmentation Performance**

We evaluated the performance for multi-organ segmentation using different starting slice positions. DSC are reported as mean ± SD to reflect performance variability. All results are detailed in Table 2.

The left kidney demonstrated the best overall performance, maintaining high DSCs across all starting positions: 0.870 ± 0.154, 0.825 ± 0.221, and 0.808 ± 0.242 for the caudal-approach, mid-approach, and cranial-approach, respectively. Notably, it was the only organ showing no statistically significant differences between any starting positions (all $p > 0.0167$).

The box plots (Figure 3) show that for all organs, the highest DSCs reached above 0.8, with some approaching or nearly reaching 1.0. However, the box plots also reveal instances of very low DSC values approaching 0 across various organs and approaches, indicating significant variability in segmentation performance.

Starting slice level had a significant impact on most organs. Organs demonstrated various patterns in segmentation performance depending on the starting level. The liver showed a significant decrease in performance as the starting position moved superiorly, with DSC dropping from 0.821 ± 0.192 with caudal-approach to 0.702 ± 0.259 with cranial-approach ($p < 0.01$). Smaller organs, such as the pancreas, adrenal glands, and gallbladder, showed the most pronounced impact of starting position. For these organs, performance significantly decreased when changing from a caudal-approach to a cranial-approach (all $p < 0.01$).

Larger organs, such as liver, kidneys, and spleen, consistently demonstrated higher DSCs compared to smaller organs across all approaches. A moderate correlation was observed across all settings when using caudal-approach ($r_s = 0.731$, $p < 0.01$), mid-approach ($r_s = 0.698$, $p < 0.01$), and cranial-approach ($r_s = 0.699$, $p < 0.01$) (12,13). As shown in Table 3, when correlation coefficients were calculated separately for each organ, fair correlations were demonstrated for almost all items, particularly in smaller organs.

We also investigated the impact of including negative prompts on segmentation performance across different organs, focusing specifically on the caudal-approach (Table 4 and Supplemental Figure 2). All organs except the liver ($p = 0.32$) and spleen ($p = 0.27$) demonstrated significant increases in DSC ($p < 0.01$) with the inclusion of negative prompts.

Next, in Figure 4, we show the highest DSC masks, excluding cases where the ground truth segmentations were incomplete. The highest-performing masks, as visualized, generated for each organ in 3D were nearly indistinguishable from the ground truth.

On the other hand, there were cases where the performance fluctuated significantly due to differences in the approach. We present an example of the liver segmentation results in Figure 5. The DSC seemingly decreased by 0.564 (from 0.924 with caudal-approach to 0.360 with cranial-approach). For the caudal-approach, the initial slice segmentation appears to have been easier due to clear contrast with the surroundings. The cranial-approach resulted in lower DSC values compared to the caudal-approach. Visual inspection of the segmentation results showed incomplete masking and potential inclusion of the inferior vena cava in the liver mask, particularly in areas where boundaries between the liver, inferior vena cava, and abdominal wall were less distinct.

## Discussion

To our knowledge, this is the first research that not only validates the performance of zero-shot SAM 2 on radiology images but also considers the impact of prompt input strategies, such as slice positioning and negative prompts. Our findings demonstrate the potential of SAM 2, a general-purpose segmentation model, in segmenting abdominal organs from CT scans. SAM2 showed promising performance for larger organs with clear boundaries, such as the liver, kidneys, and spleen, achieving mean(median) DSC of 0.821 - 0.891. Although SAM 2 was not specifically designed for medical image analysis, its notable performance suggests potential applicability to a wide range of any organ and lesion. The choice of initial prompt position had a significant impact on segmentation accuracy and the optimal position depended on the organ. Excluding negative prompts led to a significant decrease in DSC for all organs except the spleen and liver, highlighting their importance in segmentation accuracy. SAM 2 struggled with smaller and less defined structures such as the adrenal glands, pancreas, and gallbladder, resulting in lower DSCs. Interestingly, we observed a moderate positive correlation between organ volume and DSCs (Spearman's $r_s = 0.731$, $p < 0.01$), suggesting that volume size is one of several key factors influencing segmentation accuracy.

While prior studies have explored zero-shot segmentation performance in CT and MRI [16], our research extends this by investigating the impact of input prompts, potentially driving further performance enhancements when these insights are integrated with previous findings. A key advantage of SAM 2 is its ability to generate segmentation masks with just a few clicks on a single slice, drastically reducing the workload for radiologists who previously relied on labor-intensive manual annotations. Furthermore, optimizing prompt input strategies is essential for achieving even greater model performance. While the scores are lower compared to previous supervised methods, which can achieve mean DSCs in the upper 0.9 range for some organs, they are still notably high for a zero-shot prediction. Moreover, the ability to segment an entire 3D volume by simply selecting and clicking on a target structure in a single slice is particularly significant. This aligns with challenges typically observed in abdominal organ segmentation, even with supervised 3D models. Notably, supervised approaches like TotalSegmentator (based on nnUNet [17]), UNet [18], SegUNet [19], and SwinUNETR [20] also tend to show lower DSC for bilateral adrenal glands and gallbladder compared to other organs [21], a trend mirrored in SAM 2's performance. Segmentation performance can be inferred to depend on multiple factors related to the 3D morphology, volume size, and contrast with surrounding tissues of target structure. These findings suggest the importance of optimizing prompts taking into account the characteristics of the targeted structure.

Our study had several limitations. Firstly, our validation dataset was limited to abdominal CT scans, one of the most standard modalities. There are publicly available datasets such as Vertebral Segmentation [22], TotalSegmentator's MRI [23] and Duke Liver datasets[24], which include segmentation masks for various anatomical structures and modalities. Expanding our validation using these resources would allow for a more robust evaluation. Additionally, as our approach was designed to address zero-shot performance validation, we did not perform any additional training such as fine-tuning. Performance improvements can be expected by using task-specific supervised methods instead of zero-shot. Furthermore, while we used an automated approach to evaluate a large number of organs, there is potential for improved accuracy through manual prompts inputting.

In conclusion, SAM 2 has demonstrated promising zero-shot performance in segmenting certain abdominal organs in CT scans, particularly larger organs with clear boundaries, highlighting its potential for cross-domain generalization in medical imaging. However, further improvements are needed for smaller and less distinct structures. Our study underscores the importance of applying general models to unseen medical images and optimizing input prompts, which together could significantly enhance the accuracy of medical image segmentation.

## Data Availability

The data that support the findings of this study, the Total Segmentator dataset, are publicly available at Zenodo (https://zenodo.org/records/10047292) with the identifier doi:10.5281/zenodo.10047292.

**Tables:**

Table 1: Descriptive Statistics of Organ Volumes in Voxels Derived from CT Scan Mask Volumes. Min: Minimum; Max: Maximum; n: Number of samples.

| Organ | Mean | Min | Max | n |
|---|---|---|---|---|
| Liver | 465,008.6 ± 156,091.00 | 19,768 | 963,401 | 119 |
| Right Kidney | 39,381.57 ± 18,122.20 | 216 | 79,713 | 108 |
| Left Kidney | 41,246.74 ± 21,144.60 | 666 | 129,706 | 111 |
| Spleen | 71,730.34 ± 45,884.40 | 13,818 | 303,676 | 115 |
| Gallbladder | 6,247.61 ± 4,902.72 | 170 | 20,763 | 89 |
| Pancreas | 18,526.41 ± 8,502.56 | 707 | 37,855 | 116 |
| Right Adrenal Gland | 1,101.86 ± 465.47 | 216 | 2,590 | 118 |
| Left Adrenal Gland | 1,259.03 ± 522.81 | 135 | 2,977 | 115 |

Table 2: DSCs for multi-organ segmentation by different approaches (caudal, mid, and cranial). The table presents the mean DSC ± SD for each organ by different approaches. p values from Wilcoxon signed-rank tests are provided for comparisons between approaches (caudal vs mid, caudal vs cranial, and mid vs cranial). DSC: Dice Similarity Coefficient; SD: Standard Deviation.

| Organ | Approach | Mean DSC | *p* value |
|---|---|---|---|
| Liver | caudal | 0.821 ± 0.192 | caudal vs mid: < 0.01 |
| | mid | 0.754 ± 0.223 | caudal vs cranial: < 0.01 |
| | cranial | 0.702 ± 0.259 | mid vs cranial: 0.07 |
| Right Kidney | caudal | 0.862 ± 0.189 | caudal vs mid: 0.03 |
| | mid | 0.862 ± 0.212 | caudal vs cranial: 0.16 |
| | cranial | 0.801 ± 0.270 | mid vs cranial: < 0.01 |
| Left Kidney | caudal | 0.870 ± 0.154 | caudal vs mid: 0.40 |
| | mid | 0.825 ± 0.221 | caudal vs cranial: 0.15 |
| | cranial | 0.808 ± 0.242 | mid vs cranial: 0.15 |
| Spleen | caudal | 0.891 ± 0.131 | caudal vs mid: < 0.01 |
| | mid | 0.839 ± 0.187 | caudal vs cranial: 0.017 |
| | cranial | 0.768 ± 0.302 | mid vs cranial: 0.56 |
| Gallbladder | caudal | 0.527 ± 0.288 | caudal vs mid: 0.95 |
| | mid | 0.531 ± 0.291 | caudal vs cranial: < 0.01 |
| | cranial | 0.461 ± 0.314 | mid vs cranial: 0.08 |
| Pancreas | caudal | 0.353 ± 0.168 | caudal vs mid: 0.92 |
| | mid | 0.361 ± 0.197 | caudal vs cranial: < 0.01 |
| | cranial | 0.287 ± 0.209 | mid vs cranial: < 0.01 |
| Right Adrenal Gland | caudal | 0.203 ± 0.222 | caudal vs mid: < 0.01 |
| | mid | 0.177 ± 0.235 | caudal vs cranial: < 0.01 |
| | cranial | 0.112 ± 0.178 | mid vs cranial: < 0.01 |
| Left Adrenal Gland | caudal | 0.308 ± 0.234 | caudal vs mid: < 0.01 |
| | mid | 0.252 ± 0.226 | caudal vs cranial: < 0.01 |
| | cranial | 0.226 ± 0.238 | mid vs cranial: 0.08 |

Table 3: Spearman correlation coefficients between ground truth of organ volumes and DSCs in caudal, mid, and cranial levels. The table presents the $r_s$ values for each organ at different levels. $r_s$: Spearman's rank correlation coefficient.

| Organ | $r_s$ (caudal) | $r_s$ (mid) | $r_s$ (cranial) |
|---|---|---|---|
| Liver | 0.328 ($p < 0.01$) | 0.0489 ($p = 0.60$) | 0.163 ($p = 0.08$) |
| Right Kidney | 0.231 ($p = 0.0164$) | 0.347 ($p < 0.01$) | 0.509 ($p < 0.01$) |
| Left Kidney | -0.0107 ($p = 0.91$) | 0.293 ($p < 0.01$) | 0.295 ($p < 0.01$) |
| Spleen | 0.38 ($p < 0.01$) | 0.307 ($p < 0.01$) | 0.355 ($p < 0.01$) |
| Gallbladder | 0.499 ($p < 0.01$) | 0.509 ($p < 0.01$) | 0.469 ($p < 0.01$) |
| Pancreas | 0.475 ($p < 0.01$) | 0.386 ($p < 0.01$) | 0.379 ($p < 0.01$) |
| Right Adrenal Gland | 0.371 ($p < 0.01$) | 0.424 ($p < 0.01$) | 0.278 ($p < 0.01$) |
| Left Adrenal Gland | 0.452 ($p < 0.01$) | 0.339 ($p < 0.01$) | 0.447 ($p < 0.01$) |

Table 4: Comparison of multi-organ segmentation performance without negative prompts. For each organ, the table shows the mean DSC ± SD. The Difference columns represent the change when negative prompts are excluded (negative values indicate lower performance without prompts). The p value shows the results of Wilcoxon signed-rank tests comparing performance with and without negative prompts for each organ. DSC: Dice Similarity Coefficient.

| Organ | Mean DSC | Difference | *p* value |
|---|---|---|---|
| Liver | 0.785 ± 0.244 | -0.036 | 0.32 |
| Right Kidney | 0.858 ± 0.203 | -0.004 | < 0.01 |
| Left Kidney | 0.847 ± 0.192 | -0.023 | < 0.01 |
| Spleen | 0.867 ± 0.213 | -0.024 | 0.27 |
| Gallbladder | 0.438 ± 0.338 | -0.089 | < 0.01 |
| Pancreas | 0.277 ± 0.197 | -0.076 | < 0.01 |
| Right Adrenal Gland | 0.084 ± 0.151 | -0.119 | < 0.01 |
| Left Adrenal Gland | 0.190 ± 0.230 | -0.118 | < 0.01 |

## Figures

Figure 1: Flow diagram illustrating the CT scan selection process from the TotalSegmentator dataset for evaluation of SAM 2.

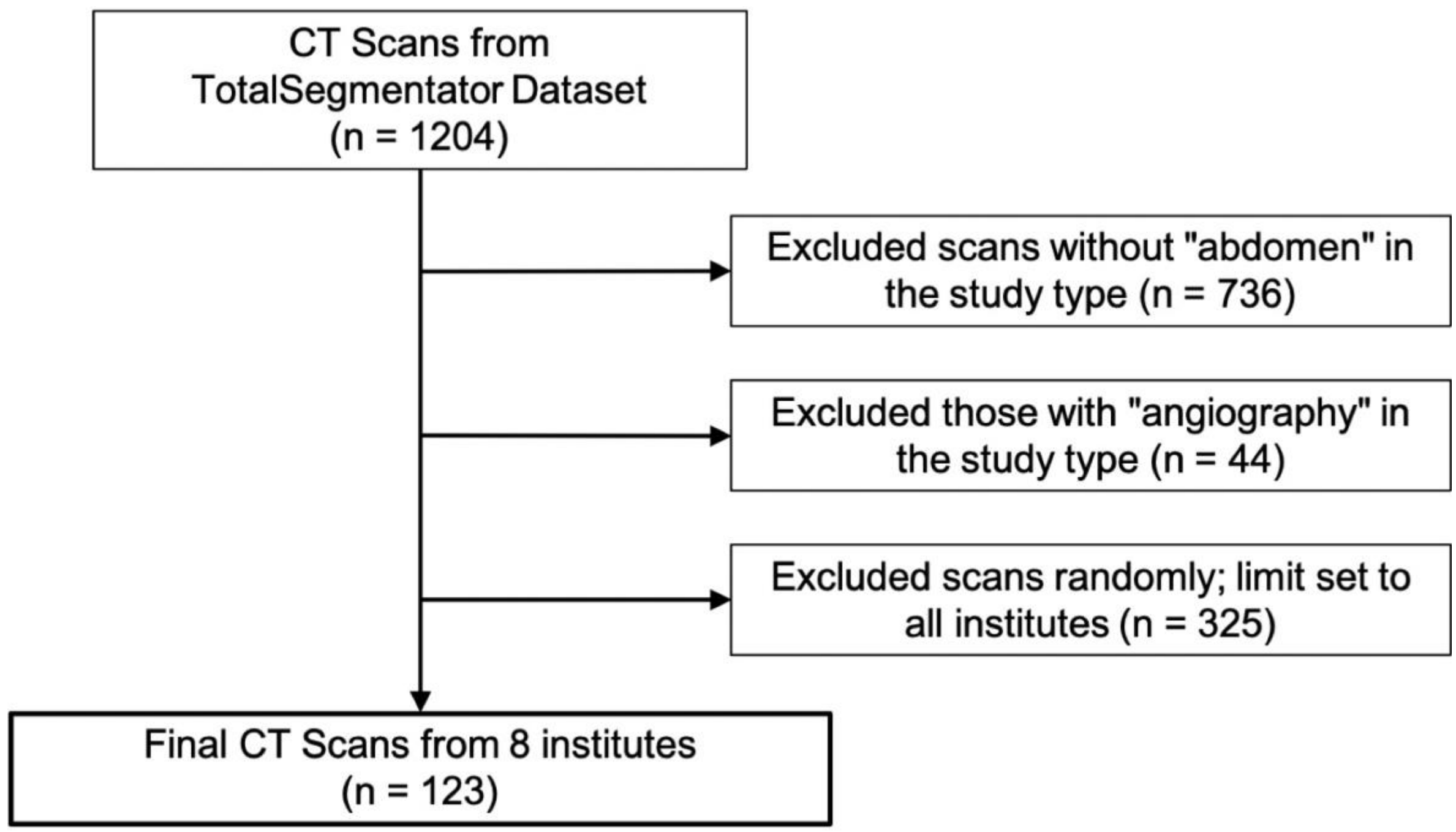

Figure 2: Workflow of 3D Medical Image Segmentation Using SAM 2. This figure illustrates a two-stage process for 3D medical image segmentation. The top row shows the Prompt Input Process, where a slice is selected from a CT volume as an initial frame, and XY-coordinates are added as prompts for organ segmentation (red crosses: positive prompts, blue crosses: negative prompts). These are then input into SAM 2. The bottom row depicts the 3D Segmentation Process, where SAM 2 performs forward and reverse predictions to generate 3D segmentation masks at both cranial and caudal sides. These masks are ultimately merged to create a full 3D segmentation mask of the target organ.

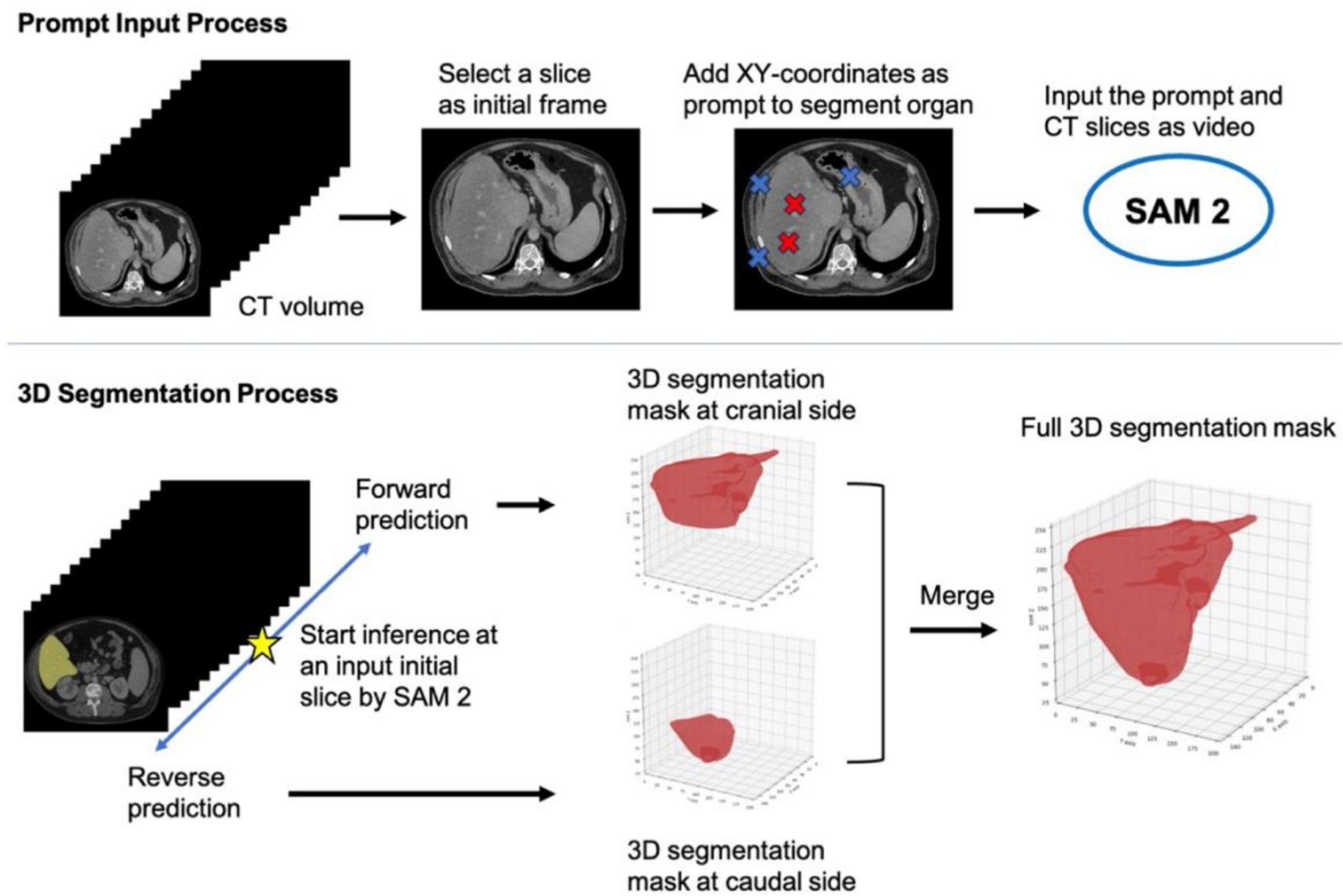

Figure 3: Box plots of DSCs for eight organs (displayed in separate subplots) across three approaches: caudal-approach, mid-approach, and cranial-approach. Each subplot shows the distribution of DSCs (y-axis, range 0-1) for a specific organ, with the three approaches compared along the x-axis. DSC: Dice Similarity Coefficient.

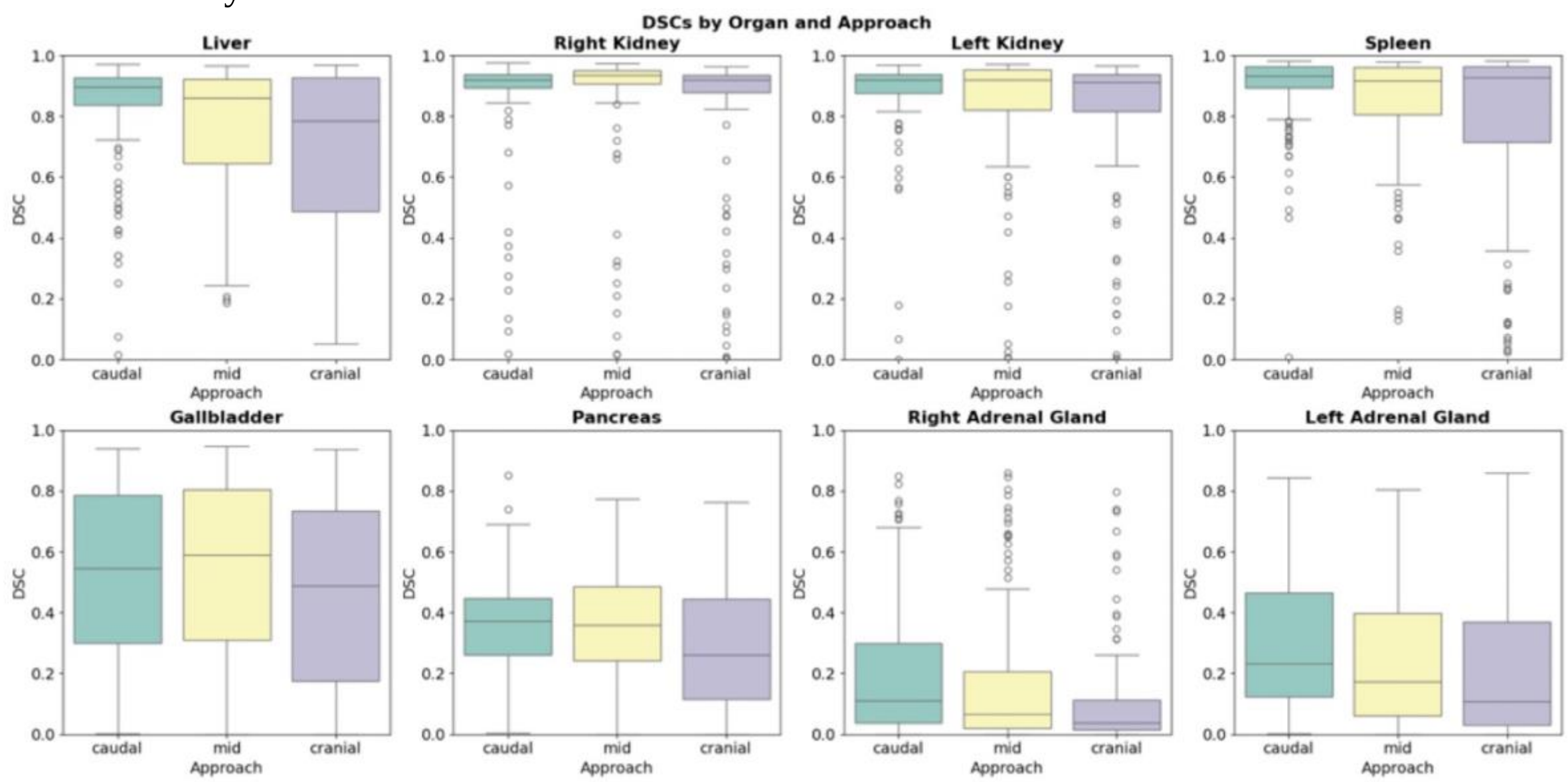

Figure 4: Successful segmentation results for eight abdominal organs. Each row shows a different organ with ground truth (left) and predicted (right) 3D masks. Values in parentheses indicate the DSC for each segmentation. DSC: Dice Similarity Coefficient.

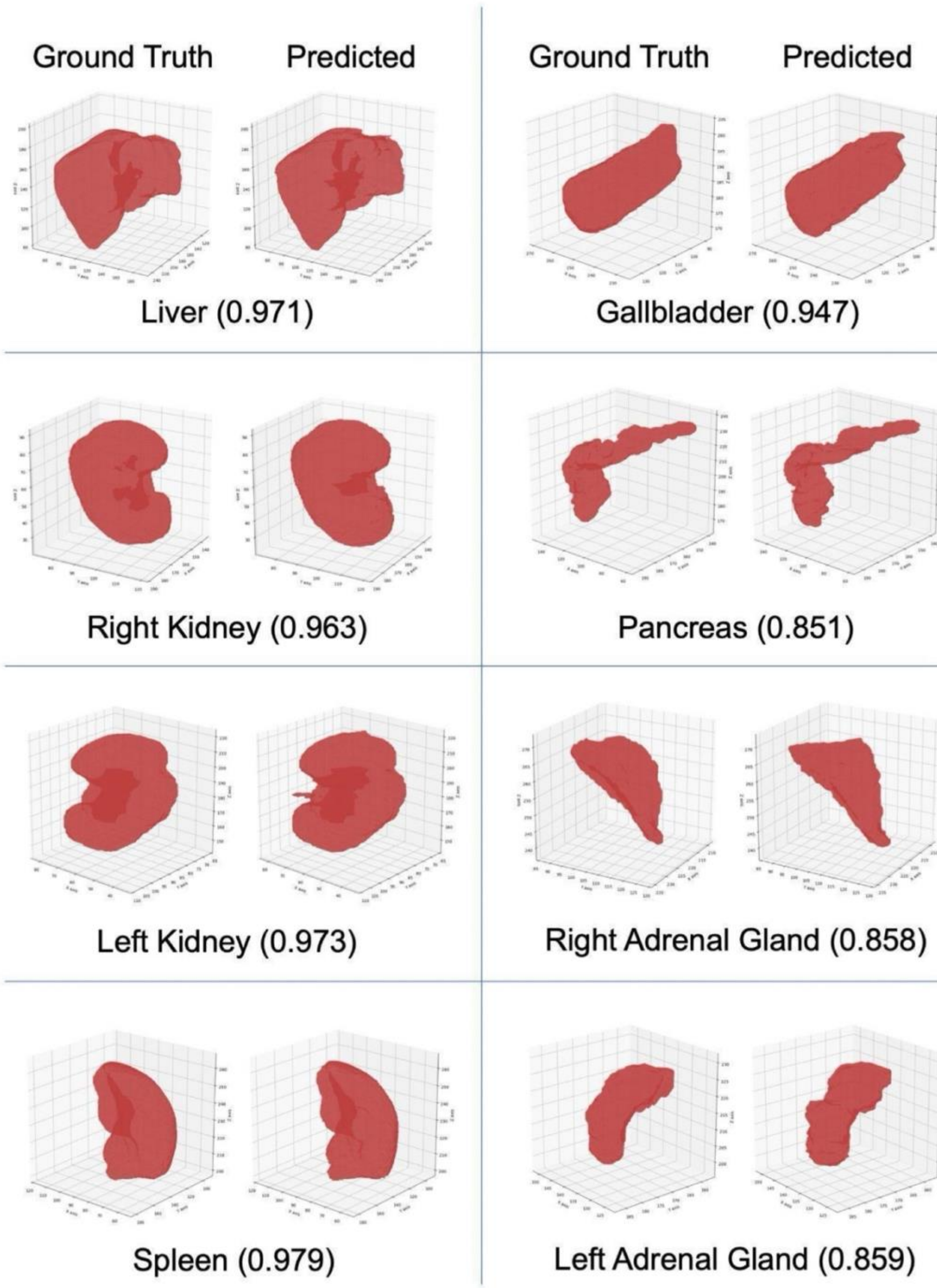

Figure 5: Comparison of segmentation results using caudal-approach and cranial-approach, showing 2D axial slices with Ground Truth and Predicted Mask for both initial slices (top row, with yellow representing the ground truth of liver, blue and red points indicating negative and positive prompts, respectively), alongside 3D renderings of liver segmentation for Ground Truth, Caudal-approach, and Cranial-approach (bottom row).

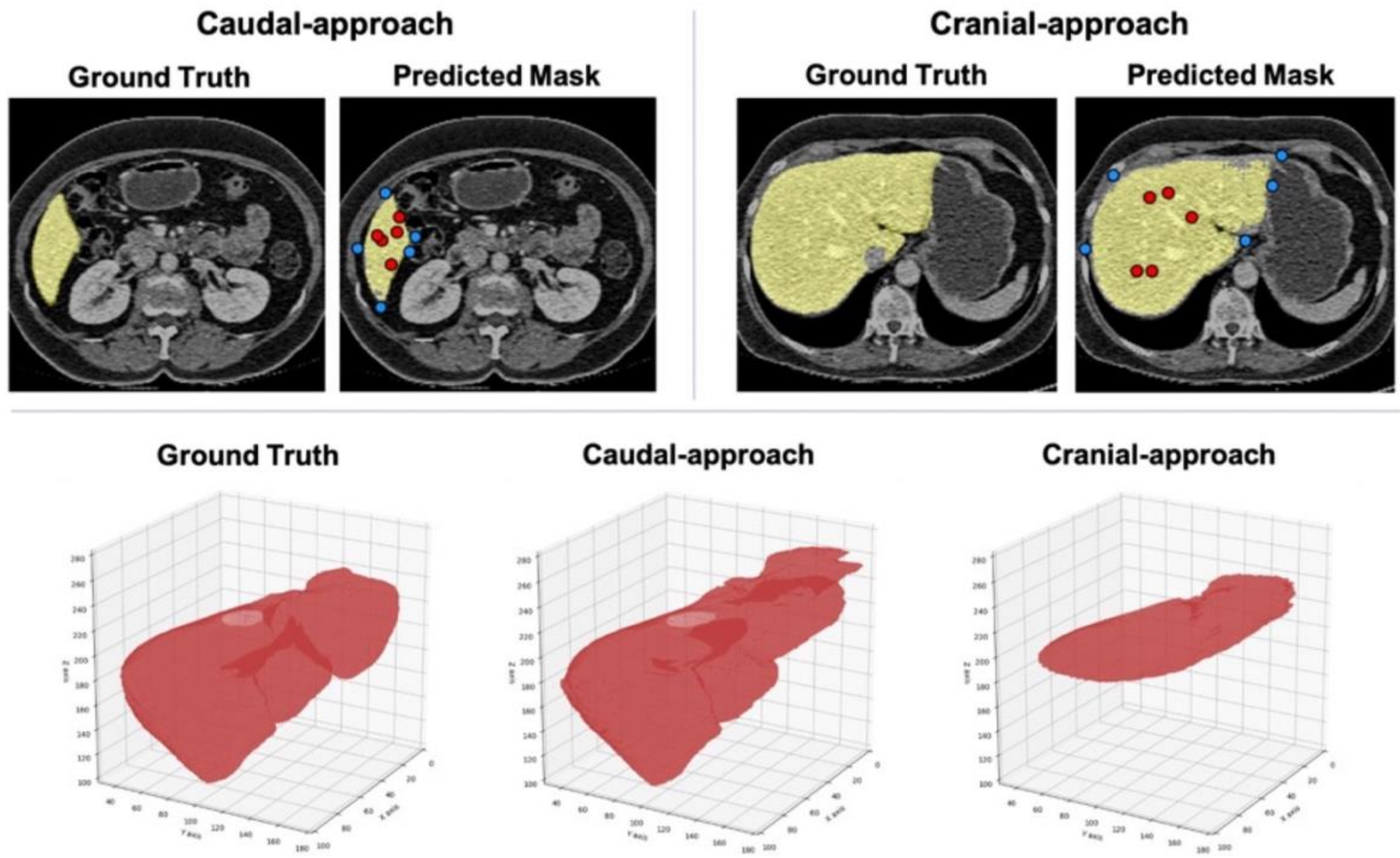

## Supplemental Materials

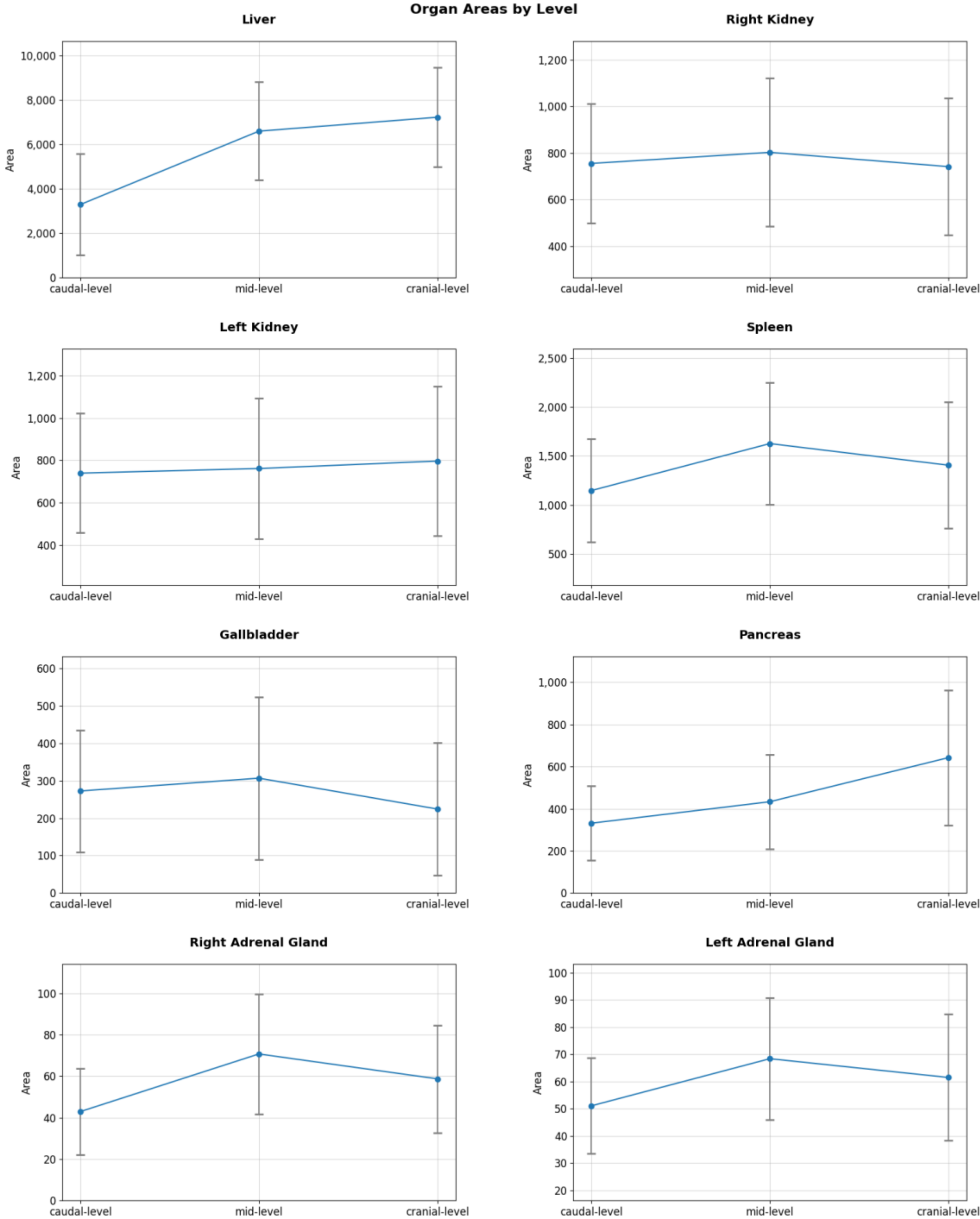

Supplemental Figure 1: Comparison of organ areas across different levels. The graph displays the mean areas (in voxel) of organs at the caudal-level, mid-level, and cranial-level. Error bars represent standard deviations.

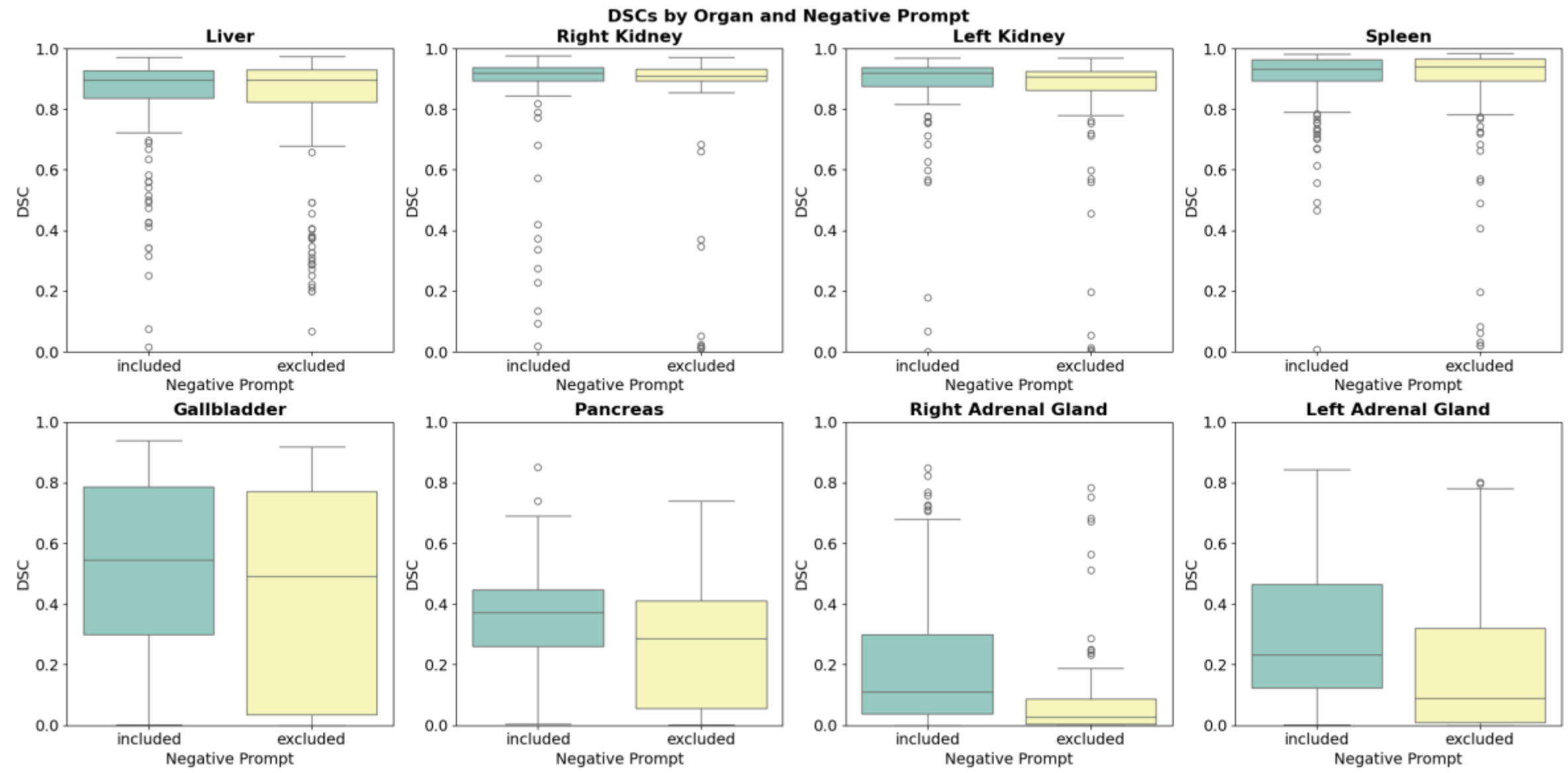

Supplemental Figure 2: Box plots comparing DSCs for eight organs with and without the inclusion of negative prompts. For each organ, DSC values are compared between two conditions: when negative prompts are included (left boxes) and when they are excluded (right boxes). DSC: Dice Similarity Coefficient.